\documentclass{PoS}
\usepackage{wrapfig}
\title{New Physics Interpretations with GAMBIT}

\ShortTitle{New Physics Interpretations with GAMBIT}

\author{\speaker{Peter Athron}\thanks{On behalf of the GAMBIT collaboration.}\\
        Monash University\\
        E-mail: \email{peter.athron@coepp.org.au}}

\abstract{I present recent results from the Global and Modular
  Beyond-the-Standard-Model Inference Tool (GAMBIT)
  collaboration. Global fits with GAMBIT have been carried out on a
  variety of models including supersymmetric models, scalar singlet
  dark matter, fermionic and vector Higgs portal dark matter and
  axions. Here I focus on a recent GAMBIT study interpreting collider
  constraints on electroweakinos (arXiv:1809.02097).  We show that
  when the neutralinos and charginos are the only light states of the
  MSSM, there are scenarios which evade LHC constraints for any mass
  of the lightest neutralino and the lightest chargino, i.e. the
  profile likelihood shows no constraint in this plane when one only
  considers the possibility of excluding new physics.  Intriguingly,
  in addition we also find that excesses in the data can lead to
  closed confidence level contours, indicating a preference for light
  neutralinos and charginos over the Standard Model. We find the
  excess has a local significance of 3.3 sigma when combining ATLAS
  and CMS 13 TeV searches, but this drops to 2.9 sigma when including
  8 TeV searches as well.}

\FullConference{
European Physical Society Conference on High Energy Physics - EPS-HEP2019 -\\
			10-17 July, 2019\\
			Ghent, Belgium}

\begin{document}
\section{GAMBIT}
The Global and Modular Beyond the Standard Model Inference Tool
(GAMBIT) is a collaboration and software tool
\cite{Athron:2017ard,Balazs:2017moi,Workgroup:2017myk,Workgroup:2017lvb,Workgroup:2017bkh,Workgroup:2017htr}
for global fits of Standard Model (SM) extensions. Global fits are the best
way to assess the collective impact of many experiments on physics
beyond the Standard Model (BSM), and is also the best way to assess
the true impact of an individual experiment.  Realistic extensions of
the SM tend to have large multi-dimensional parameter
spaces and make predictions for a wide variety of collider and
astrophysical observables. To understand the impact of searches on a
BSM physics theory one should do a global fit of the new physics
model.  To do this one should: i) combine all experimental results
using rigorous statistics to determine the composite likelihood; ii)
scan over the full parameter space with an intelligent sampling
algorithm and iii) finally to visualize the results one may project
onto planes of interest by profiling or marginalizing the composite
likelihood.  Global fitting can also be used for model comparisons, to
demonstrate which model is favored by the data.

\begin{figure*}[t]
  \centering
  \includegraphics[height=0.3\columnwidth]{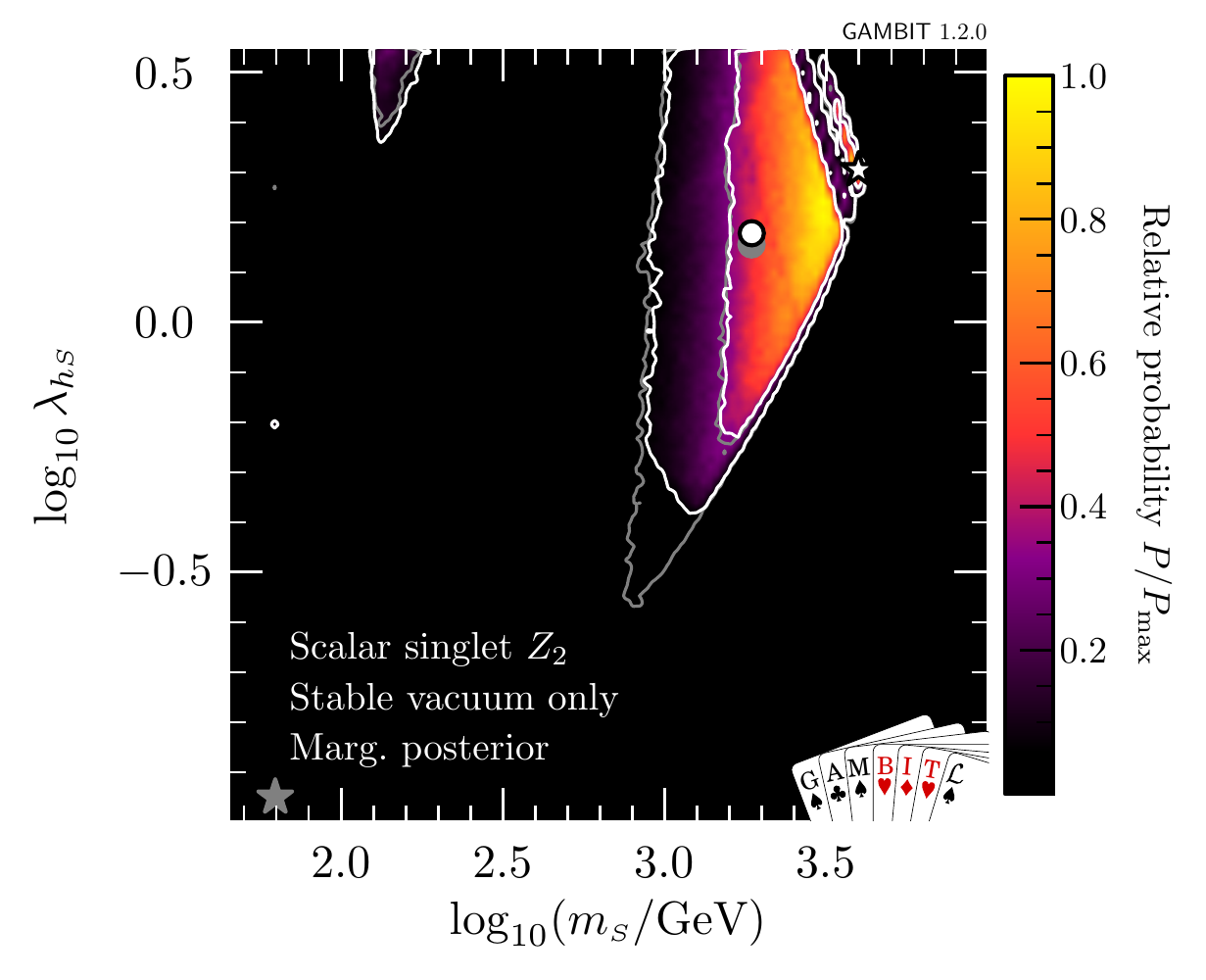}
  \includegraphics[height=0.3\columnwidth]{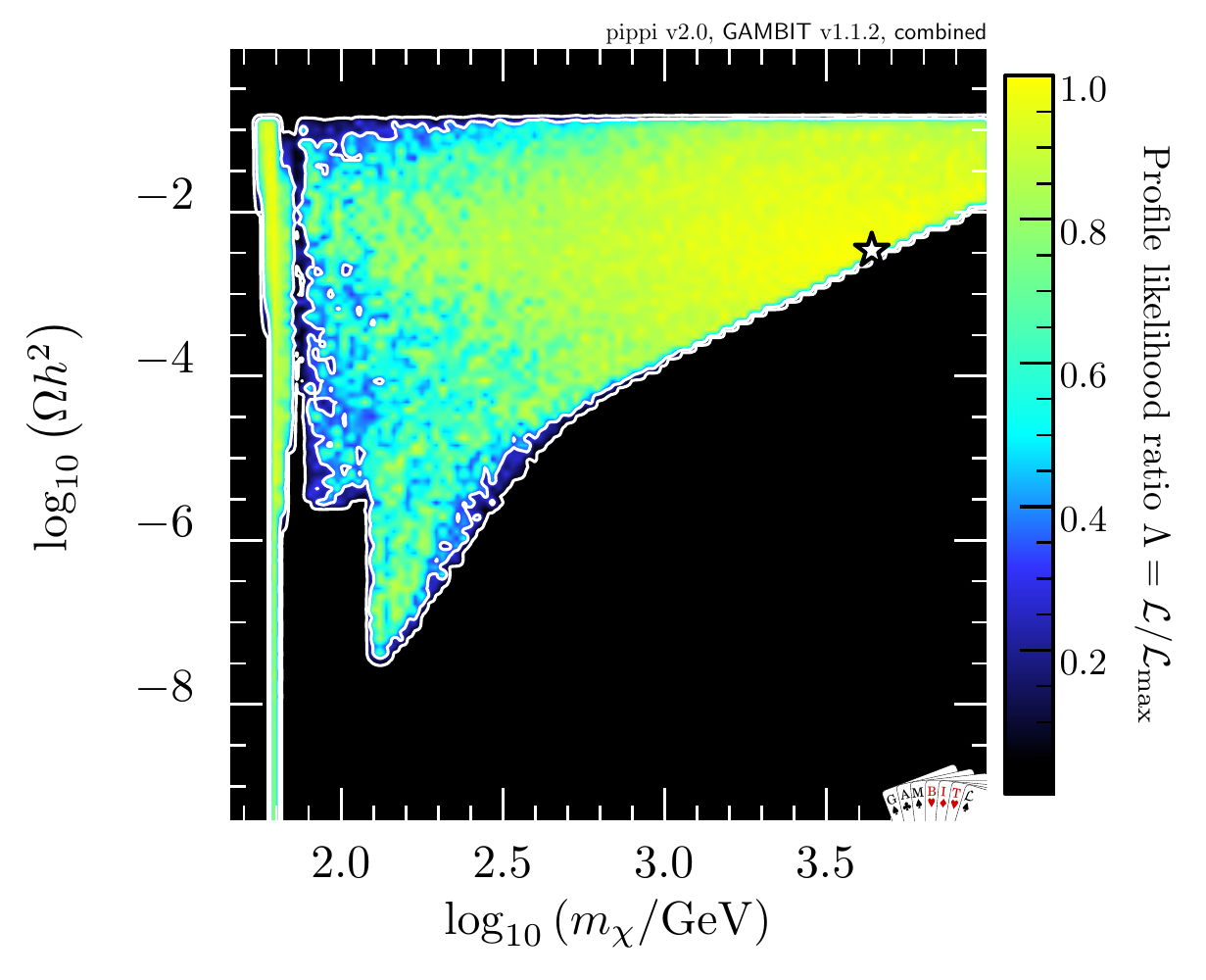} \\
  \includegraphics[height=0.3\columnwidth]{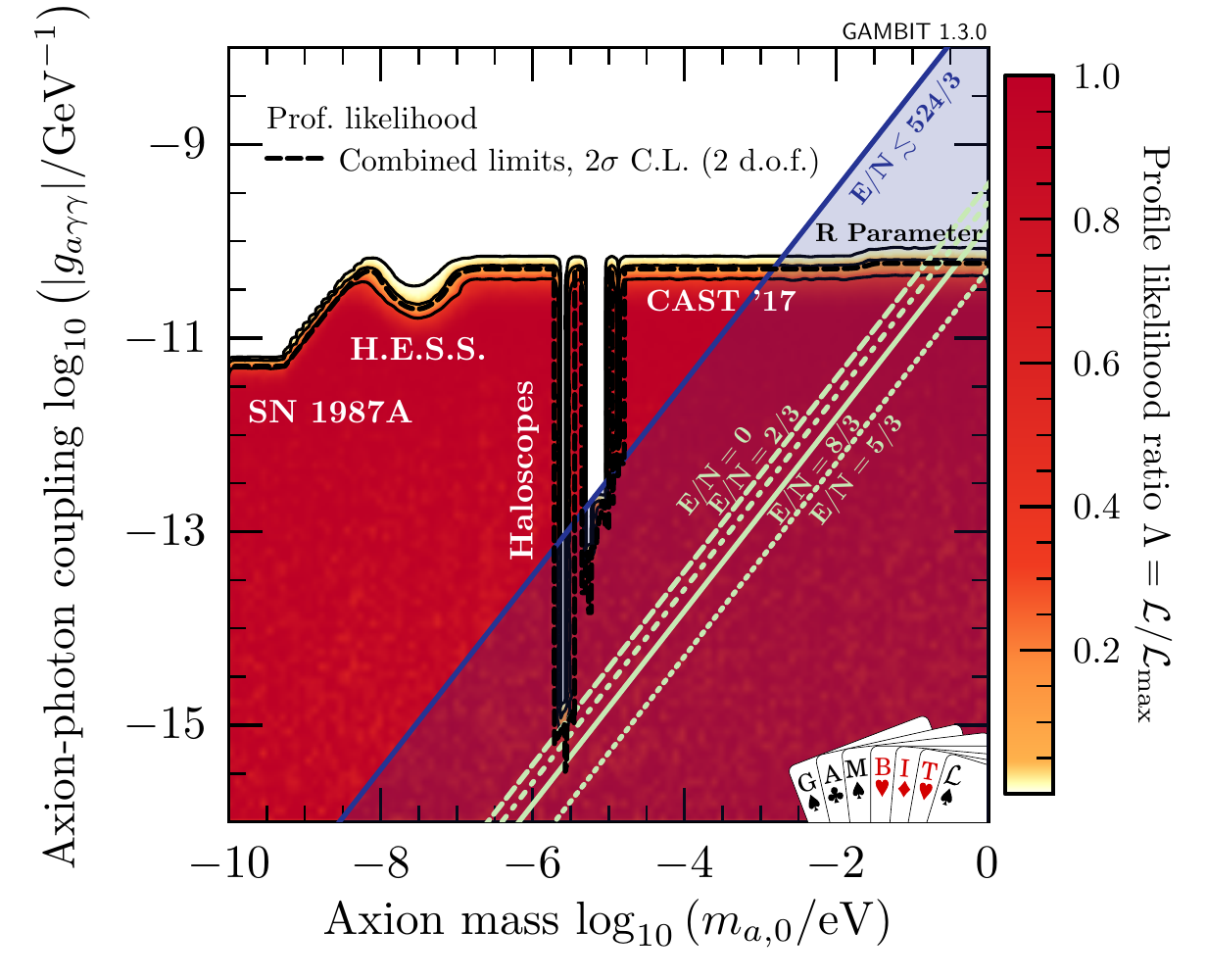}
   \includegraphics[height=0.3\columnwidth]{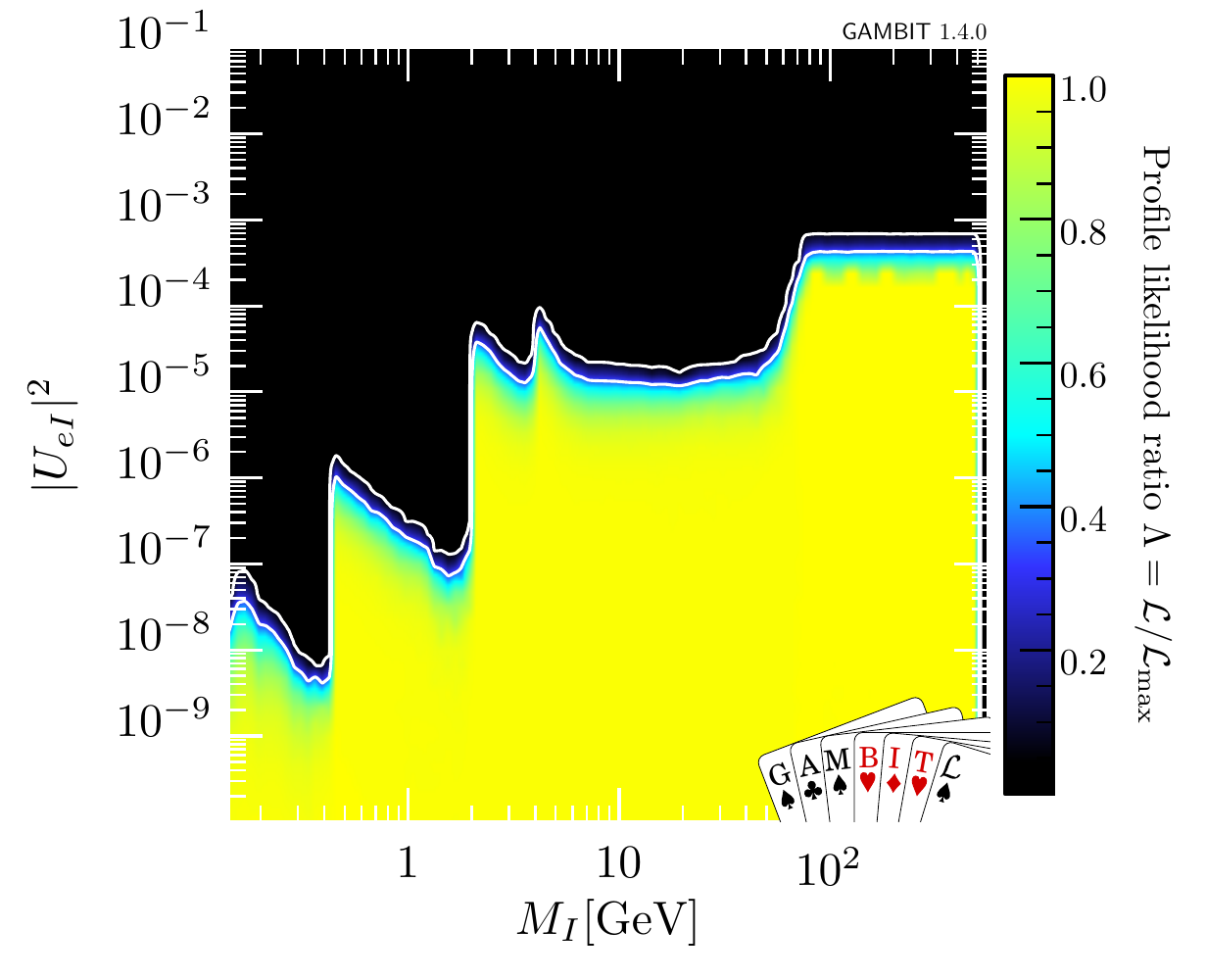}
  \caption{Results from recent papers using GAMBIT.  The top left panel shows the marginalized posterior from the scalar singlet dark matter model, taken from Ref.\ \cite{Athron:2018ipf}. The top right panel shows the profile likelihood for a Majorana fermion dark matter candidate taken from Ref.\ \cite{Athron:2018hpc}.  In the bottom left panel the profile likelihood for general axion-like particle models is shown, reproduced from Ref.\ \cite{Hoof:2018ieb}.  In the bottom right results for the Standard Model extended by right-handed neutrinos are shown, taken from Ref.\ \cite{Chrzaszcz:2019inj}.  Further details can be found in the original references.}     
\label{fig:gambit_results}
\end{figure*}

GAMBIT has been used for global fits on a wide variety of models,
including the scalar singlet dark matter model
\cite{Athron:2017kgt,Athron:2018ipf}, fermion and vector Higgs portal dark
matter \cite{Athron:2018hpc}, axions \cite{Hoof:2018ieb}, right
handed neutrinos \cite{Chrzaszcz:2019inj} and global fits of
supersymmetric extensions of the SM
\cite{Athron:2017qdc,Athron:2017yua}.  Examples of parameter estimation
from recent GAMBIT papers are shown in Fig.\ \ref{fig:gambit_results},
where one can also see that GAMBIT can be used to obtain both Bayesian
and frequentist results. GAMBIT can also be used for model comparison,
for example in Ref.\ \cite{Athron:2018hpc} the posterior odds are
calculated between each model and also compared to scalar singlet
dark matter, showing the extent to which the data discriminated
between the different models of dark matter.

Another interesting recent work has been an investigation of the
impact of collider constraints on {\it electroweakinos}
\cite{Athron:2018vxy}. In this contribution to the EPS-HEP2019
proceedings we focus on this work investigating the electroweakino sector of the the minimal supersymmetric Standard Model (MSSM). 

\section{Electroweakinos}
Electroweakinos are fermion superpartners of Higgs bosons and
electroweak gauge bosons predicted by supersymmetric extensions of the
Standard Model. These states play an important role in the naturalness
of the MSSM as the mass of Higgs superpartners (Higgsinos) can give a
tree level mass contribution to the Higgs vacuum expectation value
(VEV) that fixes the electroweak scale.  If the Higgsinos are heavier
than the weak scale then there must be cancellations between this mass
and some other parameters in order to predict the electroweak scale
found in nature.  The heavier the Higgsinos are, the greater the fine
tuning needed to enforce this cancellation.  Furthermore the lightest
neutral electroweakino (i.e.\ the lightest neutralino) provides a
plausible dark matter candidate.

Therefore we consider a model where the particles that may be light
enough to be produced at the Large Hadron Collider (LHC) are restricted to
the SM particles, including a 125 GeV SM-like Higgs boson,
the neutralinos (neutral electroweakinos) and the charginos (charged
electroweakinos).  The neutralinos are the mass eigenstates formed
from the {\it binos}, neutral {\it winos} and neutral {\it Higgsinos},
which are, respectively, superpartners of the $U(1)_Y$ gauge field,
the neutral $SU(2)$ gauge bosons, and the Higgs bosons of the MSSM. We
assume all other states predicted by the MSSM are too heavy to be
produced at the LHC.

The relevant parameters are the soft supersymmetry (SUSY) breaking bino and wino masses, $M_1$ and $M_2$ respectively, the Higgsino bilinear mass $\mu$ and the ratio of the electroweak VEVs $\tan\beta$.  $M_1$ and $\mu$ are varied in the range $[-2$ TeV,$+2$ TeV$]$, while $M_2$ is restricted to positive values in the range $[0, 2\, \textrm{TeV}]$. The range for $\tan\beta$ is $[1,70]$.




\section{Results}
First we consider a composite likelihood computed from only 13 TeV LHC
searches for electroweakinos that were available at the time of this
work and place a cap on it so that it is never greater than the
SM likelihood.  This is designed to test whether these
searches exclude realistic electroweakinos.  The complete list of
searches included can be found in Table 3 of
Ref.\ \cite{Athron:2018vxy}.

\begin{figure*}[t]
  \centering
  \includegraphics[height=0.4\columnwidth]{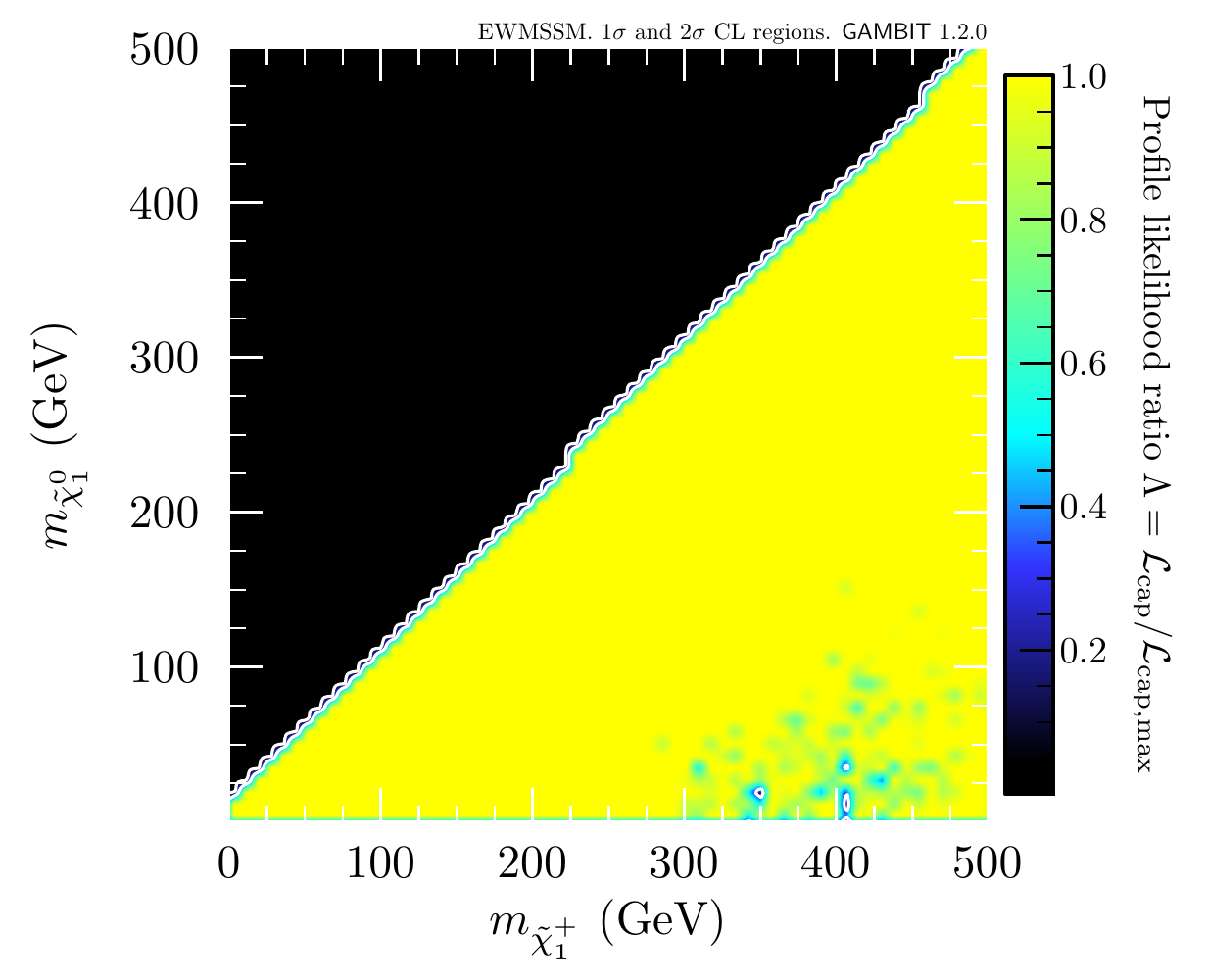}
  \caption{Capped profile likelihood from the joint likelihood for 13 TeV LHC electroweakino searches, shown in the $m_{\chi_1^0}-m_{\chi_1^\pm}$ plane.  Plot reproduced from Ref.\ \cite{Athron:2018vxy}. }     
\label{fig:capped_liklihood}
\end{figure*}

The results for this are shown in Fig.\ \ref{fig:capped_liklihood} in
the plane of the lightest neutralino ($m_{\chi_1^0}$) and the lightest
chargino ($m_{\chi_1^\pm}$).  Remarkably one can see that there is
essentially no general constraint on any point in this plane.  This
result must be interpreted carefully however.  This is {\it not}
saying that these searches are not excluding any realistic
electroweakino scenarios.  In the full four dimensional parameter
space many scenarios may be excluded.  However these results show that
there is sufficient freedom in the MSSM electroweakino sector that for
every point in the $m_{\chi_1^0}-m_{\chi_1^\pm}$ plane there is at
least one scenario on the full four dimensional parameter space that
can evade the constraints from the considered 13 TeV searches.  This
result says {\it nothing} about how much of the volume in the full
four dimensional space is constrained however.  Nonetheless it
demonstrates that the simplified model interpretations of the LHC
searches must not be over-interpreted as implying a general
constraint on this plane.  It is still possible that there are light
electroweakinos that could be discovered by the LHC or a future
experiment.

\begin{figure*}[h!]
  \centering
  \includegraphics[height=0.4\columnwidth]{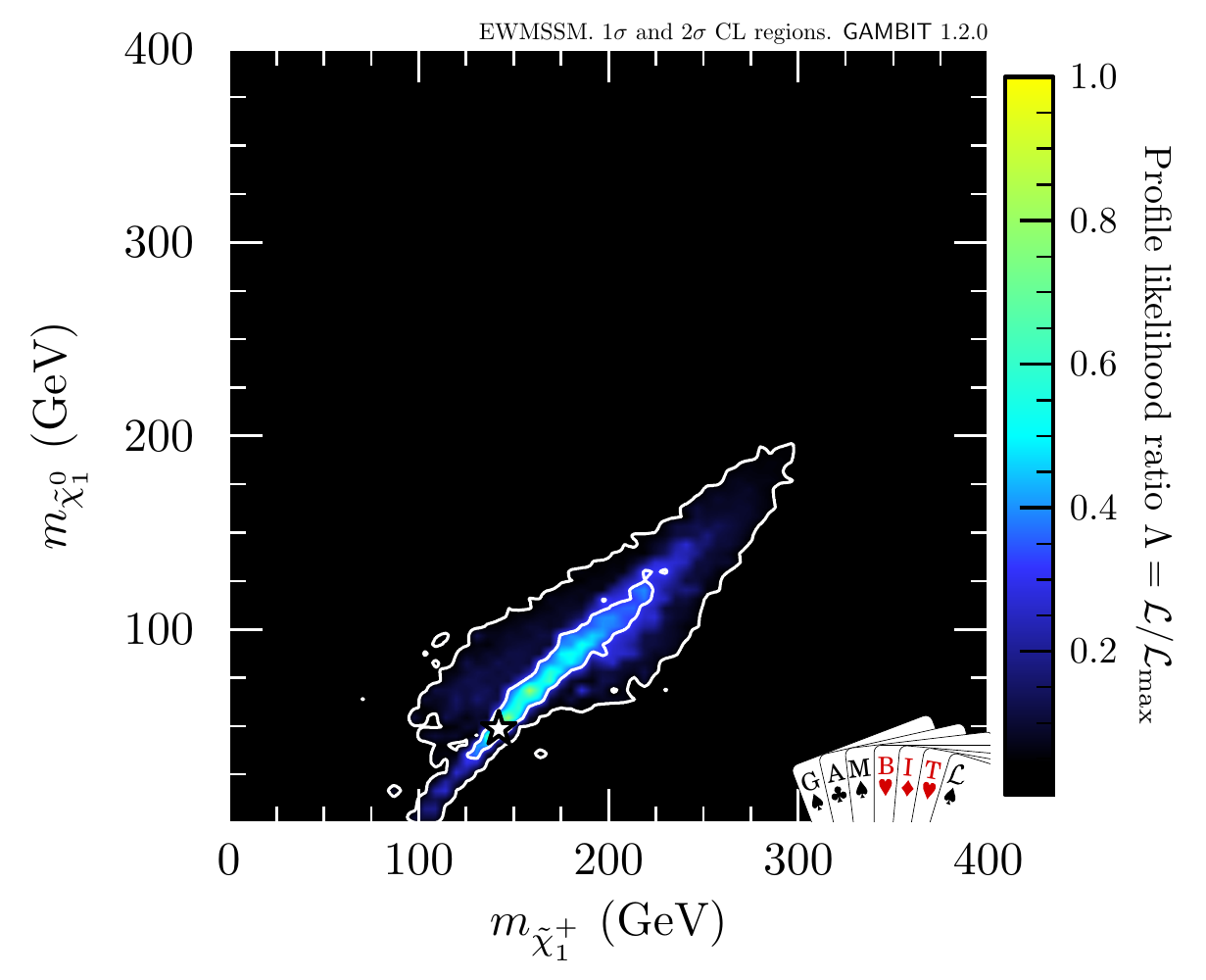}
  \includegraphics[height=0.4\columnwidth]{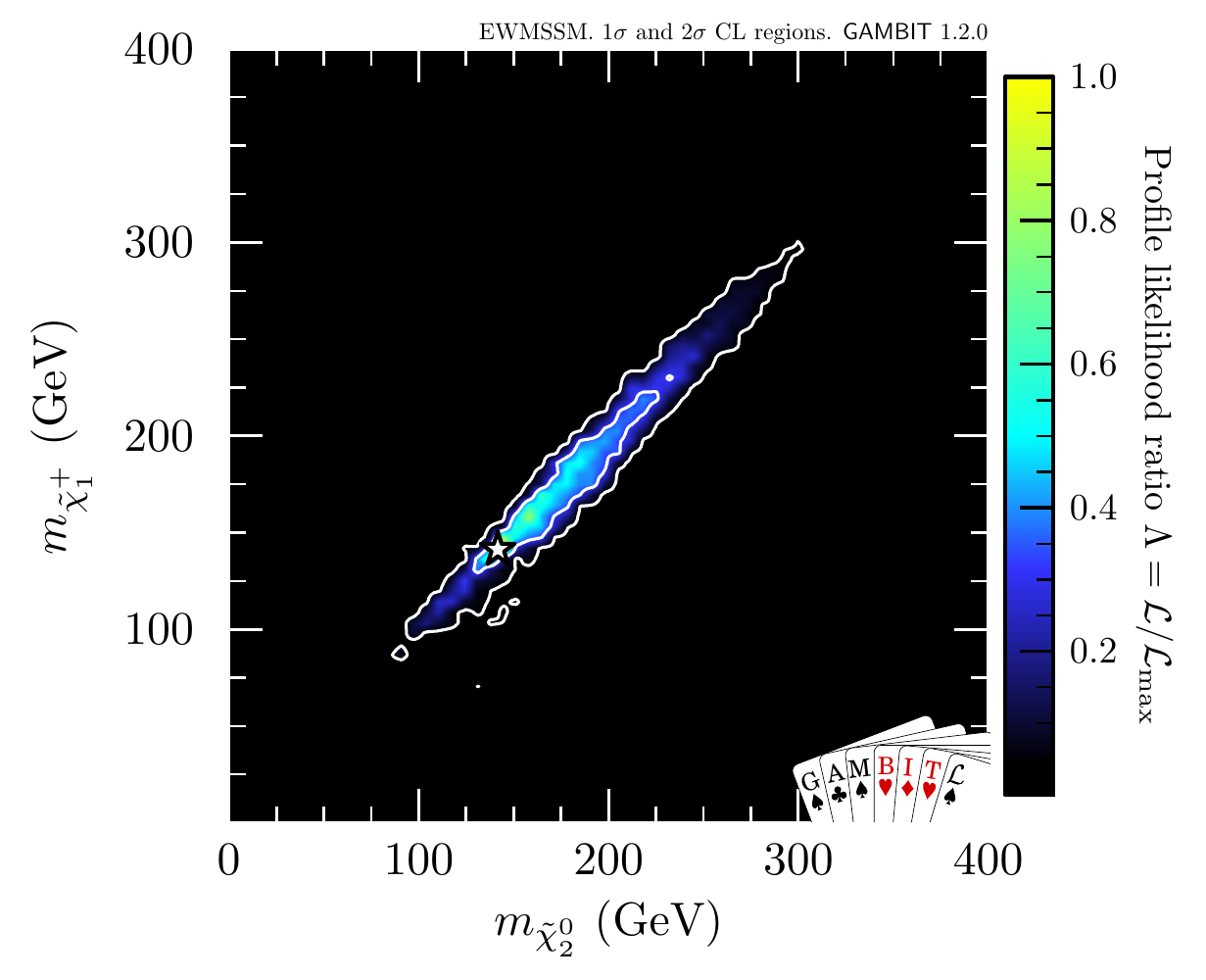}\\
  \includegraphics[height=0.4\columnwidth]{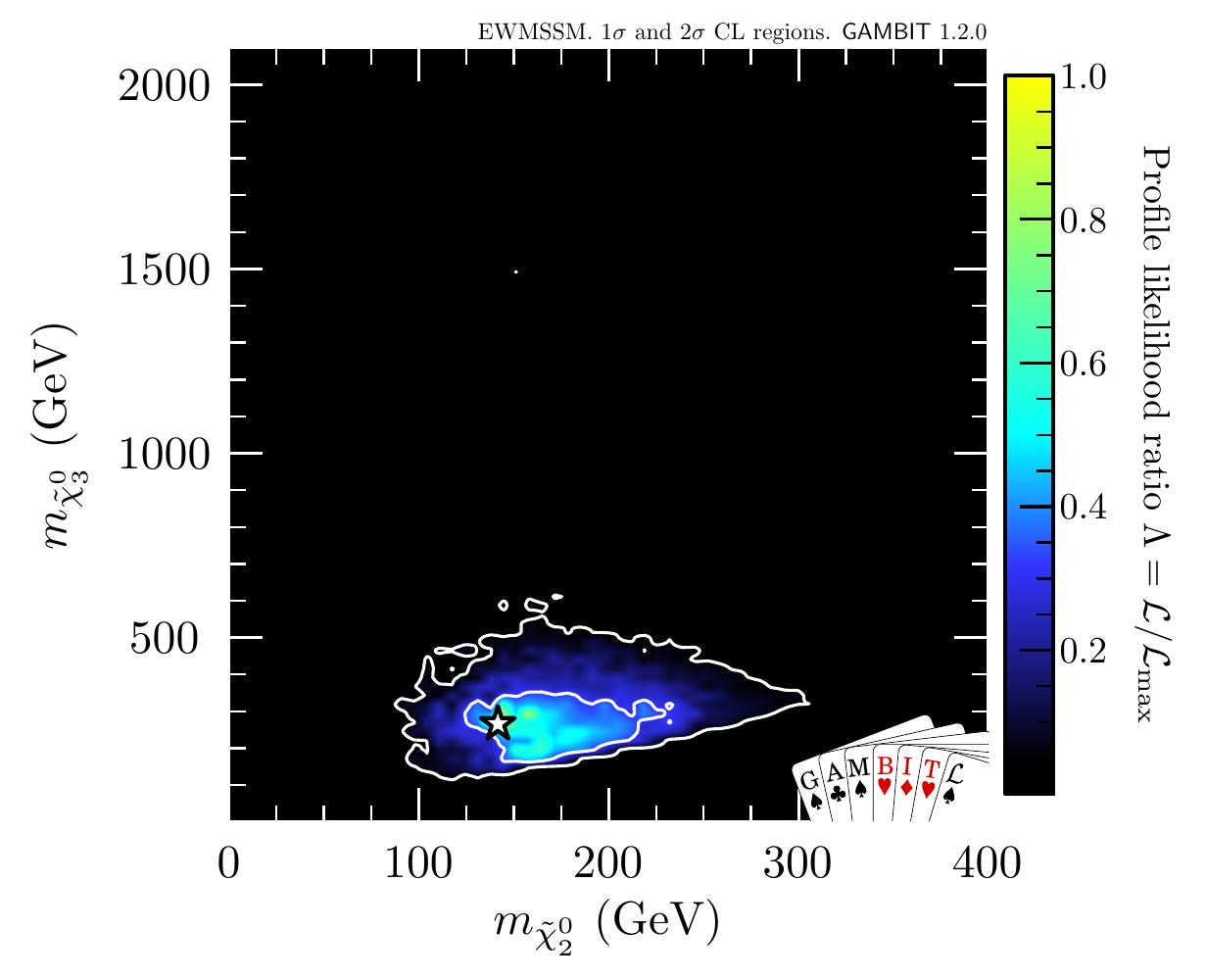}
  \includegraphics[height=0.4\columnwidth]{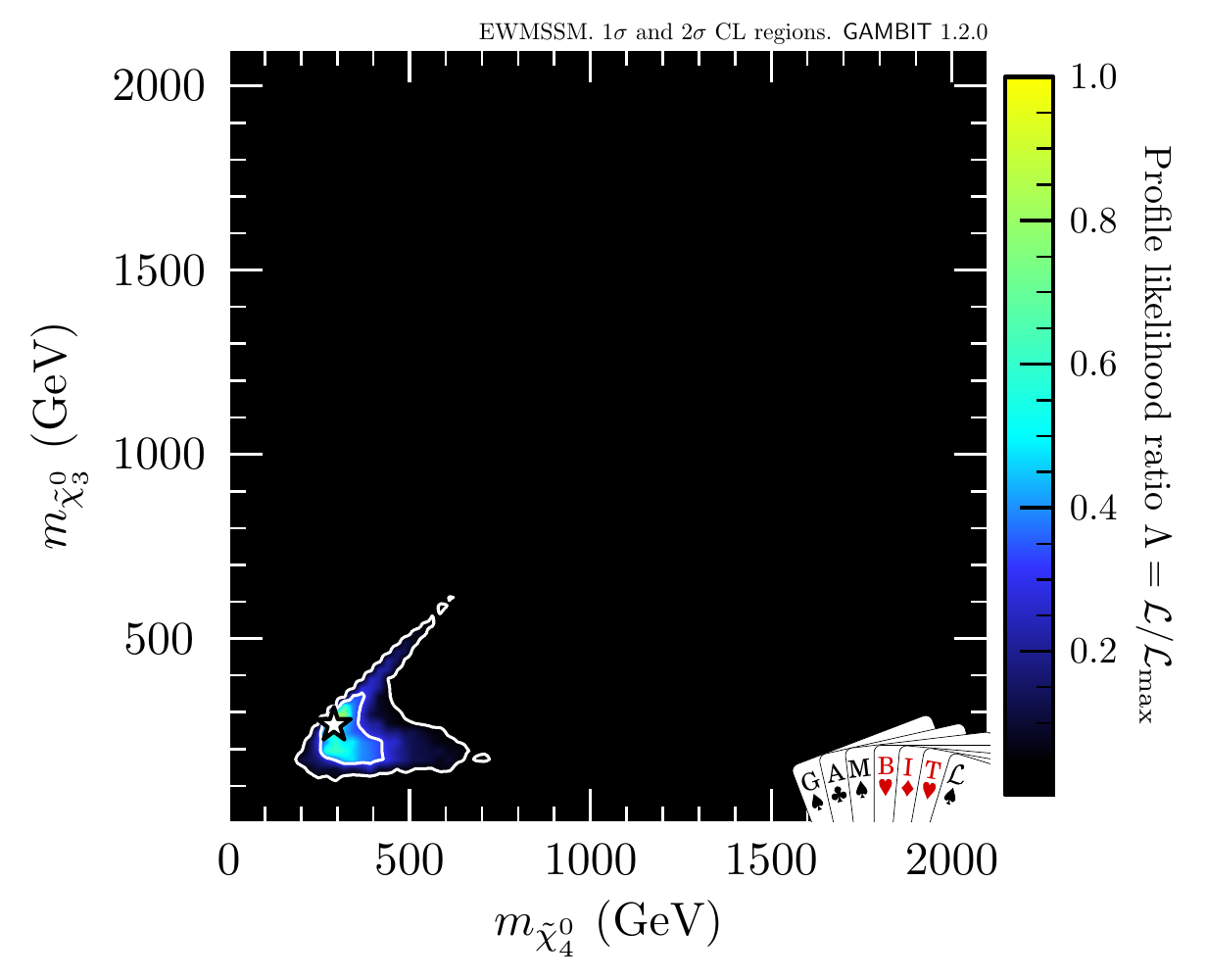}
	\caption{Profile likelihood, where the likelihood includes 13 TeV LHC electroweakino searches, LEP electroweakino searches and Higgs and Z invisible widths. Results are shown in various planes of the electroweakino masses, specifically $(m_{\tilde{\chi}_1^{\pm}},m_{\tilde{\chi}_1^0})$ (upper left), $(m_{\tilde{\chi}_2^0},m_{\tilde{\chi}_1^{\pm}})$ (upper right), $(m_{\tilde{\chi}_2^0},m_{\tilde{\chi}_3^0})$ (lower left) and $(m_{\tilde{\chi}_4^0},m_{\tilde{\chi}_3^{0}})$ (lower right). The $1\sigma$ and $2\sigma$ confidence regions are shown as contour lines and a white star is shown at the best-fit point. Plots reproduced from Ref.\ \cite{Athron:2018vxy}.
  }
  \label{fig:pole_masses_2D}
\end{figure*}

Next we remove the cap from the likelihood, so that any excesses in
the data may also be fitted, and calculate a composite likelihood
based on 13 TeV LHC electroweakino searches, LEP electroweakino
searches and Higgs and Z invisible widths.  The results are shown in
Fig.\ \ref{fig:pole_masses_2D}. Note that the in the case of a
constraint one would normally expect the low mass regions to be
black, i.e.\ ``excluded'' and the higher mass region to have larger
profile likelihood values. Instead these results show that the data actually
prefer all electroweakinos to be light, and the higher mass region (which corresponds to essentially SM-like predictions) is excluded.

As before one must be careful about the interpretation of this result
and not over-interpret it. While these plots show $1 \sigma$ and
$2 \sigma$ confidence limits, suggesting evidence for light
electroweakinos in the data, they do not provide a clear answer as to
the statistical significance of this evidence.  To assess this we
calculated a {\it local} p-value for the signal, where we used a Monte
Carlo approach to determine the distribution of our test statistic,
\begin{equation}
    q_\mathrm{LS} = -2 \log \frac{\mathcal{L}_\mathrm{joint}(\mu=1,\hat{\eta})}
                     {\mathcal{L}_\mathrm{joint}(\mu=0,\hat{\hat{\eta}})},
\end{equation}
where the numerator has the joint likelihood for analyses and the denominator has the joint likelihood for the background-only hypothesis which we construct by setting the signal to zero.

We find that the local significance is $3.3 \sigma$.  This result
includes only the 13 TeV searches, however given these results it is
clear that $8$ TeV results they are supposed to supersede can also
have an impact.  Therefore we repeated our analyses including also 8
TeV searches and this reduced the local p-value to $2.9 \sigma$.  A
second extremely important caveat here is that this local p-value does
not take account of the look else where effect.  This will reduce the
significance of the result. In any case the combined significance of
the anomalies is intriguing but smaller than other excesses that have
come and gone in recent years. 

Nonetheless a very interesting feature of our results relates to dark
matter. Since we wanted to focus only on the direct impact of the
collider searches, unlike our previous work, we did not target
explanations of dark matter in our scan by including the relic density
of dark matter or non-collider dark matter searches in our
likelihood. Indeed we assumed that all other SUSY particles are heavy
removing the possibility of sfermion co-annihilation and heavy Higgs
funnel mechanisms being present to deplete the relic density to the
observed value (or below).

\begin{figure*}[h!]
  \centering
  \includegraphics[height=0.4\columnwidth]{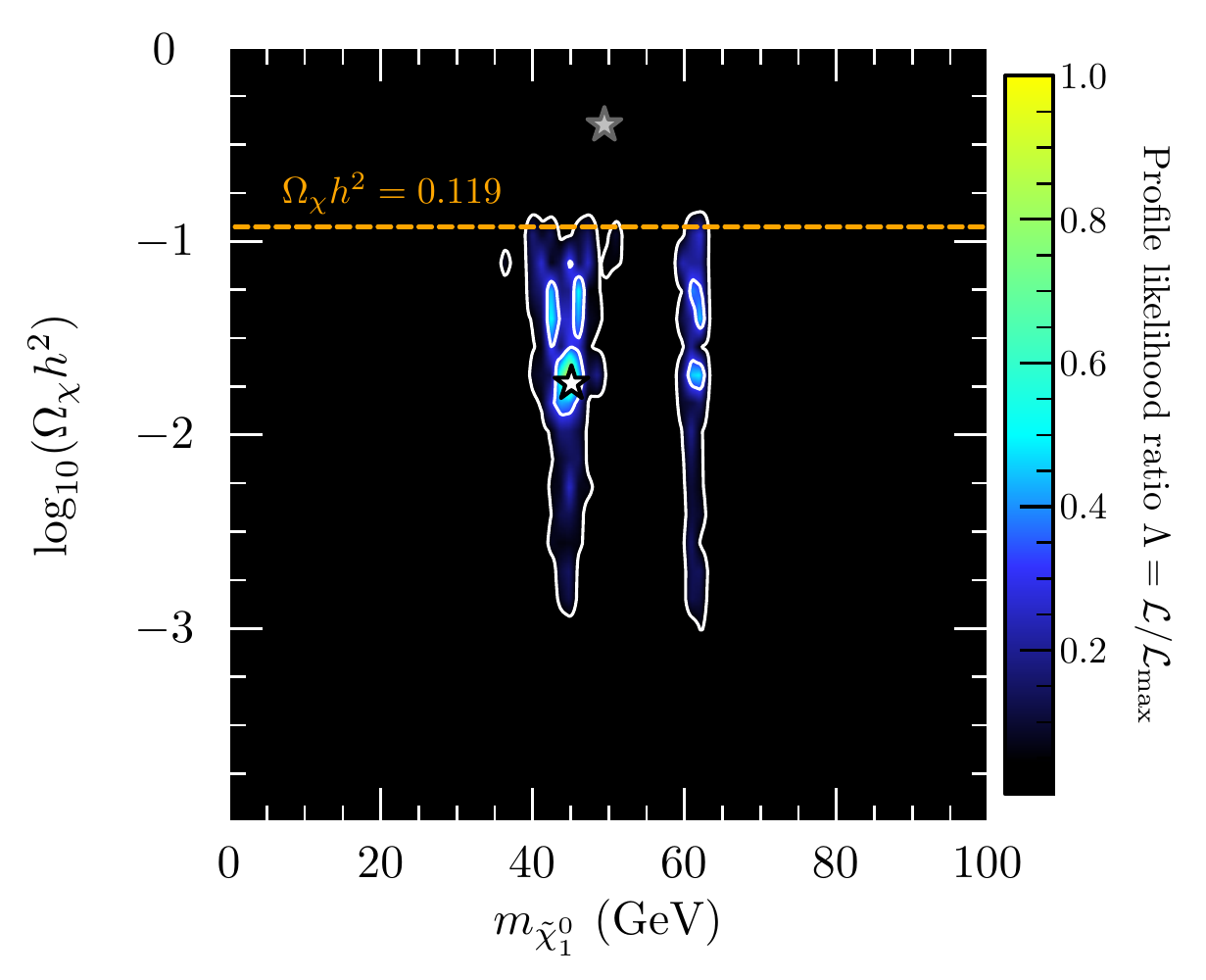}
  \includegraphics[height=0.4\columnwidth]{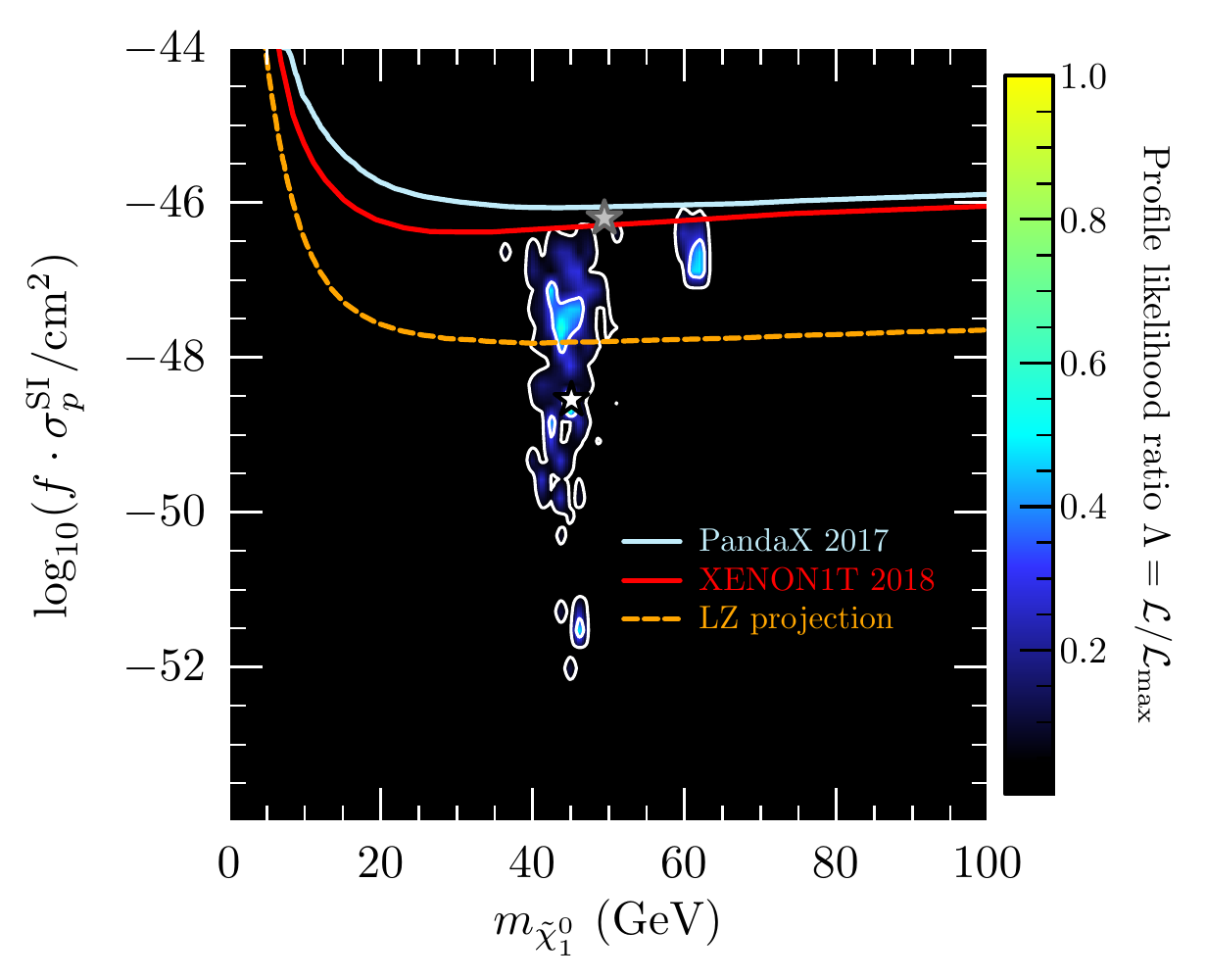}
  \caption{Profile likelihood with contributions from the LEP and LHC collider likelihoods used in the previous plot, and additional contributions from the relic density of dark matter and direct detection of dark matter.  In the left panel the results are shown for the relic density against the LSP mass, while the right panel shows the results in the plane of the spin-independent direct detection  cross-section and the LSP mass. As before $1\sigma$ and $2\sigma$ regions are shown with contour lines.  A white star marks the best fit point from the full composite likelihood, while the collider-only best-fit point is indicated by a  grey star. Plots reproduced from Ref.\ \cite{Athron:2018vxy}. 
  }
  \label{fig:relic_density_direct_detection}
\end{figure*}

However despite this the excesses automatically pushed us towards a
region of the parameter space where the relic density could be
depleted by an SM-Higgs or Z-boson funnel mechanisms.  Therefore we
post-processed our results to add likelihood contributions from the
relic density of dark matter and direct detection of dark matter.  The
results of this are shown in
Fig.\ \ref{fig:relic_density_direct_detection}.  Remarkably we find
that some of our samples can achieve a relic density at or below the
observed value and evade direct detection constraints.

\section{Conclusions}
Our results show that there is no general exclusion on light
electroweakinos from LHC. In fact there are some excesses in the data
that favor certain light electroweakino scenarios.  Furthermore a subset of
the scenarios favored by the collider searches can explain the relic
density of dark matter.  These results demonstrate that future
electroweakino searches at the LHC can be very interesting, but our
results should also be interpreted with great care.  The significance
of collider excesses is intriguing but not large enough to celebrate,
and could easily just be statistical fluctuations.  The main lesson
from this work is we must be careful how we interpret simplified model
search results. Treating them as general exclusions on non-simplified
models can exaggerate limits and could even miss statistically
significant anomalies in the data that indicate new physics!


\begin{thebibliography}{99}

\bibitem{Athron:2017ard} 
  P.~Athron {\it et al.} [GAMBIT Collaboration],
  Eur.\ Phys.\ J.\ C {\bf 77}, no. 11, 784 (2017)
  Addendum: [Eur.\ Phys.\ J.\ C {\bf 78}, no. 2, 98 (2018)]
  doi:10.1140/epjc/s10052-017-5513-2, 10.1140/epjc/s10052-017-5321-8
  [arXiv:1705.07908 [hep-ph]].



\bibitem{Balazs:2017moi} 
  C.~Bal{\'a}zs {\it et al.} [GAMBIT Collaboration],
  Eur.\ Phys.\ J.\ C {\bf 77}, no. 11, 795 (2017)
  doi:10.1140/epjc/s10052-017-5285-8
  [arXiv:1705.07919 [hep-ph]].



\bibitem{Workgroup:2017myk} 
  F.~U.~Bernlochner {\it et al.} [The GAMBIT Flavour Workgroup],
  Eur.\ Phys.\ J.\ C {\bf 77}, no. 11, 786 (2017)
  doi:10.1140/epjc/s10052-017-5157-2
  [arXiv:1705.07933 [hep-ph]].



\bibitem{Workgroup:2017lvb} 
  T.~Bringmann {\it et al.} [The GAMBIT Dark Matter Workgroup],
  Eur.\ Phys.\ J.\ C {\bf 77}, no. 12, 831 (2017)
  doi:10.1140/epjc/s10052-017-5155-4
  [arXiv:1705.07920 [hep-ph]].



\bibitem{Workgroup:2017bkh} 
  P.~Athron {\it et al.} [GAMBIT Models Workgroup],
  Eur.\ Phys.\ J.\ C {\bf 78}, no. 1, 22 (2018)
  doi:10.1140/epjc/s10052-017-5390-8
  [arXiv:1705.07936 [hep-ph]].



\bibitem{Workgroup:2017htr} 
  G.~D.~Martinez {\it et al.} [GAMBIT Collaboration],
  Eur.\ Phys.\ J.\ C {\bf 77}, no. 11, 761 (2017)
  doi:10.1140/epjc/s10052-017-5274-y
  [arXiv:1705.07959 [hep-ph]].



\bibitem{Athron:2017kgt} 
  P.~Athron {\it et al.} [GAMBIT Collaboration],
  Eur.\ Phys.\ J.\ C {\bf 77}, no. 8, 568 (2017)
  doi:10.1140/epjc/s10052-017-5113-1
  [arXiv:1705.07931 [hep-ph]].



\bibitem{Athron:2018ipf} 
  P.~Athron, J.~M.~Cornell, F.~Kahlhoefer, J.~Mckay, P.~Scott and S.~Wild,
  Eur.\ Phys.\ J.\ C {\bf 78}, no. 10, 830 (2018)
  doi:10.1140/epjc/s10052-018-6314-y
  [arXiv:1806.11281 [hep-ph]].



\bibitem{Athron:2018hpc} 
  P.~Athron {\it et al.} [GAMBIT Collaboration],
  Eur.\ Phys.\ J.\ C {\bf 79}, no. 1, 38 (2019)
  doi:10.1140/epjc/s10052-018-6513-6
  [arXiv:1808.10465 [hep-ph]].



\bibitem{Hoof:2018ieb} 
  S.~Hoof, F.~Kahlhoefer, P.~Scott, C.~Weniger and M.~White,
  JHEP {\bf 1903}, 191 (2019)
  doi:10.1007/JHEP03(2019)191
  [arXiv:1810.07192 [hep-ph]].



\bibitem{Chrzaszcz:2019inj} 
  M.~Chrzaszcz, M.~Drewes, T.~E.~Gonzalo, J.~Harz, S.~Krishnamurthy and C.~Weniger,
  arXiv:1908.02302 [hep-ph].



\bibitem{Athron:2017qdc} 
  P.~Athron {\it et al.} [GAMBIT Collaboration],
  Eur.\ Phys.\ J.\ C {\bf 77}, no. 12, 824 (2017)
  doi:10.1140/epjc/s10052-017-5167-0
  [arXiv:1705.07935 [hep-ph]].



\bibitem{Athron:2017yua} 
  P.~Athron {\it et al.} [GAMBIT Collaboration],
  Eur.\ Phys.\ J.\ C {\bf 77}, no. 12, 879 (2017)
  doi:10.1140/epjc/s10052-017-5196-8
  [arXiv:1705.07917 [hep-ph]].



\bibitem{Athron:2018vxy} 
  P.~Athron {\it et al.} [GAMBIT Collaboration],
  Eur.\ Phys.\ J.\ C {\bf 79}, no. 5, 395 (2019)
  doi:10.1140/epjc/s10052-019-6837-x
  [arXiv:1809.02097 [hep-ph]].



\end{thebibliography}



\end{document}